\documentclass{fcs}
\usepackage{graphicx}
\usepackage{epstopdf}
\usepackage{cite}
\usepackage{amsmath}
\usepackage[hyphens, allowmove]{url}
\usepackage{algorithm}
\usepackage{algorithmic}
\usepackage{array}
\usepackage[figuresright]{rotating}
\AtEndDocument{\par\leavevmode}
\makeatletter \g@addto@macro\UrlSpecials{\do\!{\newline}}\makeatother
\volumn{ }
\doi{ }
\articletype{RESEARCH~ARTICLE}
\copynote{{\copyright} Higher Education Press and Springer-Verlag Berlin Heidelberg 2018}
\ratime{Received month dd, yyyy; accepted month dd, yyyy}
\email{hejiang@ieee.org}
\title{Compiler Testing: A Systematic Literature Analysis}
\author{Yixuan TANG $^1$, Zhilei REN $^1$, Weiqiang KONG $^1$, He JIANG  $^1$ $^2$}
\address{{1\quad School of Software, Dalian University of Technology, Dalian 116000, China}\\
{2\quad School of Computer Science \& Technology, Beijing Institute of Technology, Beijing 100000, China}
}

\begin{document}
\maketitle
\setcounter{page}{1}
\setlength{\baselineskip}{14pt}

\begin{abstract}
Compilers are widely-used infrastructures in accelerating the software development, and expected to be trustworthy. In the literature, various testing technologies have been proposed to guarantee the quality of compilers. However, there remains an obstacle to comprehensively characterize and understand compiler testing. To overcome this obstacle, we propose a literature analysis framework to gain insights into the compiler testing area. First, we perform an extensive search to construct a dataset related to compiler testing papers. Then, we conduct a bibliometric analysis to analyze the productive authors, the influential papers, and the frequently tested compilers based on our dataset. Finally, we utilize association rules and collaboration networks to mine the authorships and the communities of interests among researchers and keywords. Some valuable results are reported. We find that the USA is the leading country that contains the most influential researchers and institutions. The most active keyword is ``random testing''. We also find that most researchers have broad interests within small-scale collaborators in the compiler testing area.
\end{abstract}


\section{Introduction}
\label{sec:intro}
\noindent Compilers are important infrastructure tools in software development, which provide syntax and semantics analysis for programs, as well as code optimization to accelerate software upgrades. For example, the Security Engineering group at Microsoft utilizes compilers to prioritize code review~\cite{howard2006a}; the maintenance engineers at Hewlett-Packard improve the quality of code by removing compiler diagnostics in software systems~\cite{pearse1995maintainability}.

However, compilers may also contain bugs, and in fact quite many bugs are reported for widely-used compilers such as GCC and LLVM~\cite{sun2016toward}. Buggy compilers make a source program optimized or translated into a wrong executable module, which may behave differently from the expected behavior determined by the semantics of the source program. Once this happens, it can result in disastrous software failures especially in safety-critical domains. For instance, a bug in the compiler of HAL/S had even caused the failure of the NASA Shuttle software\footnote{\url{https://history.nasa.gov/computers/Ch4-5.html}}. Even worse, developers with little knowledge about compiler bugs customarily debug the software they are developing rather than the compilers they are using, which makes compiler bugs more difficult to be found~\cite{le2014compiler,sun2016finding}. Therefore, guaranteeing the quality of compilers is a critical issue.

Compiler testing is one of the most important ways to guarantee the quality of compilers. According to the previous studies, there are three issues to be addressed: how to generate adequate test cases to test compilers, how to find the test oracles to determine whether a test case triggers bugs, and how to reduce these test cases. Furthermore, two challenges are to be addressed. First, since the inputs of compilers are complex programs with furcated syntax structures and rigorous content constraints, undefined behaviors of language specification make the first issue and the third issue be a challenge~\cite{yang2011finding}. Second, since compiler testing lacks test oracles to determine whether the outputs of compilers are semantic equivalent with the programs before they are compiled~\cite{chen2016an}, the test oracle problem makes the second issue be a challenge.

During the past decades, a great number of researchers have proposed different approaches for addressing the above issues. Some successful random test case generators have been implemented to facilitate compiler testing~\cite{lidbury2015many-core,sheridan2007practical,nagai2014reinforcing}, such as Orion~\cite{le2014compiler}, Csmith~\cite{yang2011finding,chen2013taming,regehr2012test-case}, Quest~\cite{Lindig2005Find,Lindig2005Random}, randprog~\cite{Eide2008Volatiles}, and JTT~\cite{zhao2009automated}. All of them can automatically generate abundant test programs for compilers without undefined behaviors. Simultaneously, various compiler testing techniques have been proposed to mitigate the test oracle problem, such as differential testing~\cite{sun2016finding,mckeeman1998differential}, random testing~\cite{nagai2014reinforcing,le2015randomized}, Equivalence Modulo Inputs (EMI)~\cite{le2014compiler}, mutation testing~\cite{Hariri2016Evaluating}, and metamorphic testing~\cite{tao2010an,donaldson2016metamorphic}. By employing the above testing techniques, a large number of compiler bugs can be detected. In addition, several reducers have been developed to minimize the test cases, such as Berkeley Delta\footnote{\url{http://delta.tigris.org/
}}, C-Reduce~\cite{regehr2012test-case}, and CL-Reduce~\cite{Pflanzer2016Automatic}. Thus, a set of small and valid test cases that trigger the same bugs as original ones can be reported to developers.

However, as the number of related papers increases, there are few efforts to systematically identify, analyze, and classify the influential researchers, the state-of-the-art testing technologies, the collaborations among authors, and the co-occurrence of keywords, which results in an obstacle to characterize and understand compiler testing. In this study, we employ a systematic and comprehensive literature analysis framework to overcome the obstacle. First, we perform an extensive search to identify papers related to compiler testing, and extract the most important information from papers for the consequent analysis, such as the title, the keywords, and the author(s). Then, we conduct a bibliometric analysis to identify the most influential authors and papers, as well as the widely-used compiler testing technologies, so as to present an external overview of the compiler testing area. Last, we construct three collaboration networks to analyze the communities of authors and keywords, which can present internal evidence on the influential authors and hot topics in this area.

The major contributions of this paper are summarized as follows:
\begin{itemize}
  \item We conduct a bibliometric analysis for compiler testing literature. The results show that the USA is the most influential country with a large number of excellent researchers and institutions in the compiler testing area. In addition, various types of compilers are tested, ranging from C++, Java to Pascal, whereas C compilers draw much attention from academia.
  \item We combine association rule mining and collaboration analysis to construct three networks, including the co-authorship network, the author co-keyword network, and the keyword co-occurrence network. The results show that most researchers have broad interests in the compiler testing area. These researchers distribute in several scattered communities. The keywords ``test case generation'', ``automated testing'', and ``random testing'' frequently co-occur in compiler testing.
\end{itemize}

The paper is structured as follows. Section~\ref{sec:background} illustrates the challenges and the corresponding solutions in compiler testing. We demonstrate the components of literature analysis framework in section~\ref{sec:framework}. Then, Section ~\ref{sec:results} shows the findings from bibliographic and collaboration analyses. Section~\ref{sec:related} provides an overview of related work. Section~\ref{sec:conc} concludes our paper and discusses the future direction.

\section{Background of compiler testing}
\label{sec:background}
\noindent In this section, we briefly introduce the challenges and solutions to the three issues in compiler testing.
\subsection{General compiler testing process}
\label{subsec:General compiler testing process}
\noindent Compilers can transform the source program written in high-level language into language-independent machine code, and different compilers can transform the source program into distinct binaries under various build environments~\cite{ren2018}. The process of transformation is called compilation which can be divided into three parts, i.e., frond end, middle end, and back end. In the frond end, the program can be transformed into intermediate code after the lexical analysis, syntactic analysis, and semantic analysis, in which the structure and the static semantic correctness of the program are verified. Then, in the middle end, the quality of intermediate code can be improved by machine-independent optimizers. Last, the code generator creates an executable file for the target machine according to the optimized intermediate code in the back end.

\begin{figure}[htb]
\centering
\includegraphics[scale=0.50]{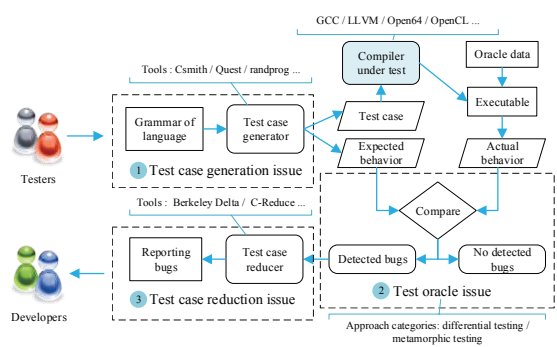}
\caption{General process of compiler testing}
%\label{fig:fragment}
\label{fig:0}
\end{figure}

In most cases, each part of transformation may contain bugs, thus comprehensive tests should be conducted to guarantee the quality of compilers~\cite{yang2011finding}. The general process of compiler testing is illustrated in Fig.~\ref{fig:0}, including three main issues. The first issue is the test case generation. The grammar of language is guided to generate test cases and the expected outputs. Several useful tools such as Quest and Csmith can randomly generate abundant test cases for testing compilers. In the second issue, test cases as inputs of the compiler under test are executed, and the actual outputs are obtained. By employing different testing methods, such as differential testing, random testing, and metamorphic testing, the actual outputs are compared against the expected outputs. For example, in differential testing, a test case can be compiled under a golden reference compiler and a test compiler. The expected output is the behavior of the golden compiler, and the actual output is generated by the test compiler with the same test case input. If there is any difference, a bug manifests in the compiler under test. The last issue is to reduce test cases which can trigger compiler bugs. Several reducers can be applied to minimize test cases, such as Berkeley Delta and C-Reduce. Once the size of a test case is small enough, the bug can be reported to developers for analyzing the test alarms and further fixing~\cite{jiang2017what}. However, each issue remains challenges that should be addressed. We present the challenges and some solutions to these challenges in the following subsections.

\subsection{Test case generation issue}
\label{subsec:test case generation issue}
\noindent We illustrate the challenge in the test case generation issue and the solutions in this subsection.

There are several commercial test suites to test the quality of compilers, such as PlumHall\footnote{\url{http://www.plumhall.com/suites.html}}, SuperTest\footnote{\url{http://www.ace.nl/compiler/supertest.html}}, GNU Compiler Collection\footnote{\url{http://gcc.gnu.org/install/test.html}}, and AC-TEST\footnote{\url{http://www.actest.co.uk/}}. Other test suites such as ACVC test suite, CppTestCase, Pascal Validation Suite, and COBOL validation tests are also employed by researchers for testing compilers. However, it is theoretically impossible to guarantee the correctness of compilers within a finite test suite. Actually, there are still many bugs in widely-used compilers, such as GCC and LLVM.

Random test case generation is an effective way to generate abundant test cases. Due to the reason that different languages are based on distinct language specifications, generating valid test cases that satisfy the corresponding language grammars is a much more difficult issue. In the case of C language, undefined behaviors make this issue a challenge. Undefined behaviors, such as zero division, signed overflow, and invalid pointer, may result in false positives. In other words, bugs are triggered by erroneous test case structures or erroneous data, rather than the compiler under test. Since possible undefined behaviors of C language may cause unexpected results and terminating execution, test cases must are free from these undefined behaviors.

The Purdom' algorithm~\cite{celentano1980compiler} is an early prominent algorithm to generate test cases based on grammar rules, and has been extended to other test case generation approaches~\cite{Boujarwah1999Testing,Chae2011An}. Then, gaussian elimination ~\cite{wu1992a} is applied to an industry example to test Fortran90D compiler. In addition, an ASM-based montages framework is proposed to generate test cases for mpC parallel programming language compiler~\cite{Kalinov2002Using}, and find a lot of inconsistent places in the Montages specifications, as well as bugs in the compiler. After that, a tool named Quest can randomly generate test cases without undefined behaviors focusing on testing the consistency of C compilers. Randprog, another random C program generator, aims at detecting bugs in compiling accesses to volatile objects. JTT, an integrated tool, is driven by test specification to automatically generate test cases for UniPhier compiler. Subsequently, Csmith extends and adapts Randprog to find bugs in C compilers, utilizing random C programs with complex control flow and data structures, such as pointers, arrays, and structs. Furthermore, CLsmith~\cite{lidbury2015many-core} has been proposed for many core compiler testing based on Csmith. However, neither Csmith nor CLsmith generates test programs for floating point test, which remains a challenge in the further test case generation.

More recently, Epiphron tools~\cite{sun2016finding} targeted compiler warning bugs support nearly all the language structures of the C language. Other semantics and skeleton equivalent test cases are generated based on metamorphic testing and Skeletal Program Enumeration (SPE)~\cite{Zhang2016Skeletal} respectively, to accelerate compiler testing. As so far, abundant test cases have been prepared to feed into compilers. Simultaneously, the expected outputs of these test cases should be collected.

\subsection{Test oracle issue}
\label{subsec:test oracle issue}
\noindent In this subsection, we illustrate the challenge in the test oracle issue and the solutions to the challenge.

Given a test case to a compiler under test and a test input to the test case, the task to distinguish the expected and correct behavior of the test case from the potential incorrect behavior is called the ``test oracle problem''~\cite{barr2015the}. However, the challenge is that it is difficult to determine whether the observed behavior is correct, because the expected behavior is difficult to be accurately described. In the literature, several approaches have been proposed to mitigate this issue. We categorize these approaches into two groups, namely the differential testing and the metamorphic testing.

Differential testing needs two or more compilers under the same specification to determine whether there is a bug by comparing the behaviors of these compilers given the same test cases as inputs. There are three strategies to implement differential testing, i.e., cross-compiler strategy~\cite{sheridan2007practical}, cross-optimization strategy~\cite{le2015randomized}, and cross-version strategy~\cite{sun2016finding}. Cross-compiler strategy detects bugs by comparing the behaviors produced by different compilers; cross-optimization strategy compares the behaviors of different optimizations implemented in a single compiler, whereas cross-version strategy uses different versions of a single compiler to determine whether there is a bug. However, to the best of our knowledge, there are only a few formal verification compilers that can be used as a golden reference compiler for testing compilers, because of the difficulty of formal verification problem~\cite{leroy2009formal,kong2015facilitating}. As a result, differential testing has its weakness when new programming languages are involved.

Notably, metamorphic testing introduces an alternative view on differential testing. If the behaviors of a set of semantically equivalent test cases dissatisfy the metamorphic relations, there is a bug manifests in the compiler under test. The advantages of metamorphic testing are that the approach can not only mitigate the test oracle problem, but also can be regarded as an effective complement to differential testing, especially when there are no available reference compilers. Furthermore, Equivalence Modulo Input (EMI) which is derived from metamorphic testing adopts the equivalence relation under a set of oracle data as the metamorphic relation. The key insight behind EMI is to compare the results of source test case and its equivalent variants under the same oracle data to determine whether there is a bug in a compiler. Any detected deviant behavior on the same oracle data indicates a bug in the compiler. In fact, EMI has three instantiations, i.e., Orion, Athena~\cite{Le2015Finding}, and Hermes~\cite{Sun2016FindingC}. Orion stochastically prunes program statements in dead regions. Athena utilizes Markov Chain Monte Carlo optimization to guide both code deletions and insertions in dead regions, and Hermes allows mutations in both live and dead regions to help more thoroughly stress test compilers. An empirical study~\cite{chen2016an} shows that different testing approaches are effective at detecting distinct types of compiler bugs. Cross-optimization strategy is more effective at detecting optimization-related bugs, and cross-compiler strategy can substitute EMI and Cross-optimization strategy in detecting optimization-irrelevant bugs. It is time consuming to test software, test case prioritization is a challenging task to accelerate software testing~\cite{meihong2012}, especially in compiler testing~\cite{Chen2017Learning}.

\subsection{Test case reduction issue}
\label{subsec:test case reduction issue}
\noindent  We present the challenge in the test case reduction issue and the corresponding solutions in this subsection.

To report a compiler bug, a test case that triggers the bug must be as small as possible because it is more difficult to reproduce due to the lengthy bug reports with diverse sentences and large size of test case~\cite{lixiaochen2018}. In most cases, test cases are manually reduced which is laborious and time-consuming. Automatic test case reduction is required to help minimize test cases before reporting them to compiler developers. However, in the case of C language, undefined behaviors make this issue a challenge, because the test case should be free from undefined behaviors during the reduction process, and the reduced test case must trigger the same bug as the original one.

There are several reducers to automatically reduce test cases, including Berkeley Delta, C-Reduce, and CL-reduce. Berkeley Delta is based on delta debugging algorithm which reduces test cases at line granularity. C-Reduce is a state-of-the-art tool for reducing C programs which refers to abundant static and dynamic analyses to avoid undefined behaviors. Subsequently, C-Reduce is extended to CL-reduce which provides test case reduction for OpenCL kernels. Another approach adopts top-down minimization and bottom-up minimization algorithms alternately to reduce a tree structure constructed by arithmetic expressions until there is no space to minimize any more~\cite{Nagai2012Random}. As a result, a test case with thousands of lines of code can be reduced to a few lines. However, all these reduction approaches only support single-file program reduction, whereas multiple-file programs reduction and real-world projects reduction still require further efforts.

\textbf{Conclusion.} The compiler testing area includes three crucial issues, i.e., the test case generation issue, the test oracle issue, and the test case reduction issue. In order to address these three issues, two challenges need to be avoided, i.e., the undefined behaviors in test cases and the test oracle problem. In the literature, several approaches and tools are proposed to address these challenges. In order to investigate which approaches and tools are frequently employed when testing compilers, we conduct a bibliometric analysis, and present the results in Section ~\ref{sec:results}.
\section{Framework}
\label{sec:framework}

\begin{figure*}[!ht]
\centering
\includegraphics[scale=0.60]{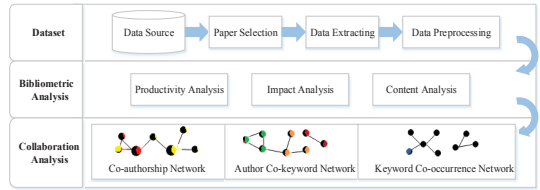}
\caption{The components of framework}
\label{fig:2}
%\label{fig:fscore}
\end{figure*}

\noindent The whole framework consists of three components, i.e., the dataset, the bibliometric analysis, and the collaboration analysis, as shown in Fig.~\ref{fig:2}. First, we construct a dataset containing the most important information of papers related to compiler testing in the dataset component. Then, the bibliometric analysis component provides three modules to present an overview of compiler testing. Last, we constructs three networks in the collaboration analysis component to present the internal evidence on collaborations between researchers and their interests. We detail each component of the framework in the following subsections.
\subsection{Dataset}
\label{subsec:dataset}
\noindent To construct the dataset, we refer to the processes of review study to find relevant published papers in journals and conference proceedings. We search three major online academic search engines, i.e., IEEE Xplore\footnote{\url{http://ieeexplore.ieee.org}}, ISI Web of Science (WoS)\footnote{\url{http://apps.webofknowledge.com}}, and Scopus\footnote{\url{https://www.scopus.com}}. These search engines are widely accepted in review studies~\cite{garousi2013a,kanewala2014testing}, and support advanced search. Then, we define a search string ``compiler \textbf{AND} (test \textbf{OR} bug)'', and limit the search within titles, abstracts, and keywords for paper selection. We do not limit a specific published time or journal/conference when conducting the searching. Therefore, the papers in our initial dataset are published before February 2018.

Since the focus of this paper is on compiler testing, many papers that target compiler verification and other software testing are included in our searching results. Thus, it is necessary to define comprehensive inclusion/exclusion criteria to select only the papers that provide evidence supporting for compiler testing.

For the inclusion criteria, we include the:
\begin{itemize}
  \item Research papers that describe at least one compiler testing technology.
  \item Cases studies and surveys of compiler testing experiences.
  \item Papers of reference lists that are relevant to compiler testing.
\end{itemize}

For the exclusion criteria, we exclude the:
\begin{itemize}
  \item Papers that are not published in English.
  \item Resources of papers that are not available online.
  \item Short papers that are less than four pages.
  \item Papers that are duplications.
  \item Papers that are not related to compiler testing.
\end{itemize}

\begin{figure}[htb]
\centering
\includegraphics[scale=0.45]{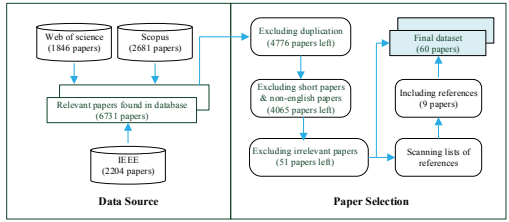}
\caption{The process of paper section}
%\label{fig:fragment}
\label{fig:3}
\end{figure}

With the above search string, we find 6,731 papers in our initial dataset. We conduct the paper selection process, and present the collection of the number of papers after performing each criterion in parentheses as shown in Fig.~\ref{fig:3}. First, we check their titles to remove duplicates, and obtain 4,776 papers. Second, we excluded those papers that are less than four pages, and are not written in English. After applying this step, 711 papers are filtered. Then, we check the titles, keywords, and abstracts to eliminate irrelevant papers. In other words, only a paper describing the solutions to at least one issue in the compiler testing area is included in our dataset. We find that most papers are filtered out in this step because these papers are related to compiler verification or other software testing process. It is time-consuming and laborious work to exclude irrelevant papers. Nonetheless, we design and conduct such a concise search string to describe the compiler testing area and include many more papers that may be related to this area in the initial dataset. Manually checking on the papers can ensure that most papers related to compiler testing are included in our dataset, and filter out those papers that do not focus on compiler testing issues. Thus, only 51 papers are left in our dataset after this step. Last, we apply the same selection criteria to the reference lists of the selected 51 papers to find additional papers. Nine papers that are not retrieved by the search keywords are included. Finally, we obtain 60 papers related to compiler testing for the following procedures.

We design a data extraction form to collect needed information to support the bibliometric analysis and the collaboration analysis, as shown in Table~\ref{table:1}. In addition to the bibliographic information of title, keywords, abstract, author(s), institution(s), country, and published year, the data form also includes the citation number of each paper which is collected from Google Scholar\footnote{\url{http://scholar.google.com}}, the identified subject of compiler under test, the tools and the methods used for test case generation, and the types of compiler testing technologies.

\begin{table}[h]
\begin{footnotesize}
\caption{Extraction data item and description}
\label{table:1}
%\label{table:features}
\centering
%\scalebox{0.55}{
\begin{tabular}{p{3.2cm}p{4.5cm}}
  \hline\noalign{\smallskip}
  Data Item & Description\\
  \noalign{\smallskip} \hline
  Title & Title of paper \\
  Author & Authors'name of paper \\
  Abstract & Abstract of paper \\
  Keywords & Keywords presented on paper \\
  Institution & Institution of author \\
  Country & Country of author \\
  Published year & Year that the paper was published \\
  Citation & Citation number of paper \\
  Subject & Types of compiler under test \\
  Data generation & Tools/methods proposed to generate test case \\
  Compiler testing technology & Types of testing method used \\
  \hline
\end{tabular}
\end{footnotesize}
%}
\end{table}

When we collect the bibliographic information from papers, we find that not all of the selected papers contain keywords due to the different formatting template of different journals/conferences. To accurately analyze the keywords, we furnish keywords information of these papers by extracting three keywords from the abstract information using the TextRank~\cite{mihalcea2004textrank} algorithm, which is a graph-based ranking model for text processing, and has been successfully used in natural language applications for term identification~\cite{Balcerzak2014Application,Rahman2015TextRank}. We select at least three keywords by the TextRank algorithm for each paper in the following analyses.

However, the items of subject, data generation, and compiler testing technology cannot be directly extracted from papers. For these pieces of information, we employ three postgraduates of Dalian University of Technology to manually identify the relevant items. Each of them needs to scan each paper to answer the following questions:

\begin{itemize}
  \item What types of compilers are tested in the paper?
  \item How the test cases are generated for compiler testing?
  \item Which testing technology is employed when testing compilers?
\end{itemize}

We adopt the most consistent answers for each question. If there are no consistent answers to a question, we invite another three postgraduates to answer the question until there is a consistent agreement.

All the needed information of selected papers in the data extraction form is constructed into our dataset. Subsequently, we conduct bibliometric analysis and collaboration analysis based on this dataset, and detail these analyses in the following subsections.

\subsection{Bibliometric analysis}
\label{subsec:bibliometric}
\noindent The bibliometric analysis consists of three modules, i.e., the productivity analysis, the impact analysis, and the content analysis. We show the details of each module in the following subsections.

\subsubsection{Productivity analysis}
\noindent The productivity analysis is mainly used to identify the most productive authors, institutions, countries, and popular topics in the compiler testing area. Thus, we calculate the number of papers for each author, institution, and country to identify the most productive ones. In order to avoid the ambiguity of the authors with the same name, we calculate the published number of each author with the institution when the paper is published. Once there are authors with the same name but different institutions, we check the homepage of authors to distinguish them. In addition, if the authors of a paper are from different institutions and counties, we calculate the distinct institution and country for once.

Then, we count the frequencies of keywords to identify the most popular topic and the trends of several popular topics. Notably, we delete the keywords ``compiler testing'' and ``compiler bugs'' when calculating the frequency of keywords, since the keywords are our search strings and the focus of this study.

\subsubsection{Impact Analysis}
\noindent The impact analysis is used to identify the influential authors and papers in the compiler testing area. We detail the measurement of the impact of papers and authors as follows.

\vspace*{2mm}
1)   Impact of papers: The motivation behind this indicator is that, the higher the citation number is, the higher impact of a paper receives. We use Google Scholar to find all papers' citation number before February $3^{rd}$, 2018. However, the newly published papers tend to have a smaller citation number compared with the previous ones. Therefore, we use Normalized Citation Impact Index (NCII)~\cite{Holsapple2015Business} which considers the impact of a publication's longevity to solve this issue. The score of NCII can be calculated as follows:
\begin{equation}
NCII=\tfrac{\mbox{\textit{Total citation per referenced publication}}}{\mbox{\textit{Publication Longevity(inyears)}}}.
\label{eq:ncii}
\end{equation}

Publication longevity indicates the number of years that a paper has been in print. With respect to this paper, the year 2018 is considered as the end point of the period.

\vspace*{2mm}
2)   Impact of authors: We utilize individual contributions of papers to measure the impact of authors. Specifically, we employ Adjusted Citation Score (ACS)~\cite{Mcclure1998Foundations}, to calculate the individual contributions based on both papers' number of the author and the citation number of each paper.

Given a set of papers \textit{P=$\left \{ p_{1},...p_{n} \right \}$} and a set of published numbers \textit{N=$\left \{ n_{1},...n_{n} \right \}$}, each paper \textit{$p_{i}$} in \textit{P} has been published by the corresponding \textit{$n_{i}$} authors in \textit{N} in our dataset. Then, the score of ACS is defined as follow:

\begin{equation}
ACS=\sum_{p \in P}\tfrac{NCII}{n}.
\label{eq:acs}
\end{equation}

We modify the calculation of ACS, and replace the citation number of each paper with the score of NCII. Thus, equation~\ref{eq:acs} evaluates a paper's quality by the corresponding NCII value.

\subsubsection{Content analysis}
\noindent The content analysis is used to identify the frequently used compilers, popular test case generators, and testing technologies. Thus, we analyze the frequency of each compiler under test, the widely-used test case generator and the test suite, the compiler testing technology, and the approach based on the manual extraction data items in our dataset.

\subsection{Collaboration Analysis}
\label{subsec:collaboration}
\noindent The collaboration analysis is mainly used to reveal the cooperative relationships between authors and their interests. Thus, we generate three collaboration networks, i.e., the co-authorship network, the author co-keyword network, and the keyword co-occurrence network, to realize this analysis. We construct these networks because the collaborations between authors can be directly reflected in the co-authorship network, the common interests among authors can be found in the author co-keyword network, and the core topics in compiler testing can be detected by similar keywords in the keyword co-occurrence network. We also employ community detecting algorithm~\cite{blondel2008fast} to find different communities in networks. In addition, all networks are visualized as undirect graphs, because the collaborations among authors and keywords can be undisputedly viewed as parallel.

\subsubsection{Collaboration networks associations}
\noindent The information of authors and keywords are needed to construct the networks. In the co-authorship network and the author co-keyword network, the nodes stand for authors, while the nodes in the keyword co-occurrence network stand for keywords. Specifically, we use association rule mining~\cite{agrawal1994fast} to help mine useful collaboration associations.

As we are interested in constructing collaboration networks, we need to identify the frequent pairs of collaborations between authors and keywords. In the co-authorship network, a pair of authors is a frequent item if the proportion of the number of papers that are co-authored by this pair of authors is above the minimal support threshold \textit{$t_{s}$}. Similarly, in the author co-keyword network, if the proportion of the number of papers that are organized using the same keywords by a pair of authors is above the minimal support threshold, we incorporate this pair of authors into frequent items. In the keyword co-occurrence network, a pair of keywords is a frequent item if the proportion of the number of papers that are organized with this pair of keywords is above the minimal support threshold. An association rule is generated from such pair if the confidence of this rule is above the minimal confidence threshold \textit{$t_{c}$}. The confidence threshold is calculated as the proportion of the number of papers that contain the frequent pair of collaborations compared with the number of papers that contain only the first one in the frequent pair.

Given the mined association rules, we can construct three collaboration networks. Each network is an undirected graph \textit{$N=\left \{ A/K,E,W \right \}$}, where the node set \textit{$A/K$} contains authors or keywords that appear in the association rules. The link set \textit{$E$} contains undirect links that connect two authors or two keywords, and the weight set \textit{$W$} represents the confident attribute indicating the strength of association rules.

\subsubsection{Community detection}
\noindent A network can consist of a large number of authors or keywords, as well as links between them. In graph theory, a node would be tightly linked with other relevant nodes, but loosely linked with irrelevant nodes. A set of highly correlated nodes is referred to as a community in the network. For example, in the author co-keyword network, authors with the same interests are most likely to be a community, because most of them focus on a specific topic in compiler testing. We use the Louvain method~\cite{blondel2008fast} implemented in the Gephi~\cite{bastian2009gephi} tool to detect communities in the networks. The Louvain method partitions each network into a finite number of communities by using an iterative modularity maximization method, rather than requiring users to specify the number of communities. The modularity is defined as follow:

\begin{equation}
\label{eq:modurity}
Q=\tfrac{1}{2m}\sum_{ij}\left [  V_{ij}-\tfrac{d_{i}d_{j}}{2m} \right ]\delta (c_{i},c_{j}) ,
\end{equation}
where the \textit{$\delta$}-function is 1 if nodes \textit{i} and \textit{j} belong to the same community, otherwise the \textit{$\delta$}-function is 0. Also, the \textit{$V_{ij}$} is 1 if the two nodes \textit{i} and \textit{j} are linked, otherwise the \textit{$V_{ij}$} is 0. \textit{m} indicates the number of links in the network, and the \textit{$d_{i}$} represents the degree number of the node \textit{i}. Each node must be assigned to a specific community. Intuitively, the links in the same community will enhance the density of the network, and perform a positive effect to increase the modularity, whereas the links across different communities have a negative effect on modularity.

\subsubsection{Visualizing the networks}
\noindent We use the Gephi~\cite{bastian2009gephi} tool to visualize the collaboration networks. Forceatlas2 layout~\cite{jacomy2014forceatlas2} is used to achieve spatialization, because this layout is convenient to investigate different communities. Nodes and links in the same community are shown in the same color, whereas the nodes and links are shown with different or similar colors in different communities. The size of a node (author/keyword) represents the number of collaborations. The larger a node is, the more authors or keywords collaborate with the node. The thickness of links represents the strength of associations rules. The wider a link is, the more times that the two nodes collaborate with each other. However, the length of links bears no meaning in this paper due to the use of Forceatlas2 layout.

\section{Results and Analysis}
\label{sec:results}
\noindent In this section, we present the results of the analyses based on our framework using the constructing dataset. In particular, we investigate the following research questions:

\begin{description}
  \item [RQ1.] What are the influential authors, institutions, and the trends in the area of compiler testing?
  \begin{description}
  \item [RQ1.1] What are the productive authors, institutions or countries?
  \item [RQ1.2] What are the frequent keywords and the trends of popular topics?
  \item [RQ1.3] What are the influential authors and papers in the area of compiler testing?
  \end{description}
  \item [RQ2.] What are the research situations of compiler testing?
  \begin{description}
  \item [RQ2.1] What compilers are frequently tested?
  \item [RQ2.2] What test cases and testing technologies are employed when testing compilers?
  \item [RQ2.3] How to reduce the large test cases before reporting?
  \end{description}
  \item [RQ3.] What are the author communities and topic communities in the compiler testing area?
  \begin{description}
  \item [RQ3.1] What are the relationships among authors of compiler testing?
  \item [RQ3.2] What are the same interests of authors?
  \item [RQ3.3] What are the frequent co-occurrence keywords in the area of compiler testing?
  \end{description}
\end{description}

We conduct the bibliometric analysis to help mine infrastructural information of compiler testing to address the former two main questions. Then, we conduct the collaboration analysis to explore the relationships among authors and keywords to address the last main question. In addition, we visualize the collaboration networks to characterize the collaborations more clearly.

\subsection{Investigation to RQ1}
\label{subsec:investigationRQ1}
\noindent We detect a large number of excellent authors and institutions that plays major roles in the development of the compiler testing area by conducting the productivity analysis and the impact analysis. In the following subsections, we only list some top-ranked results due to space restrictions.

\begin{figure}[htb]
\centering
\includegraphics[scale=0.80]{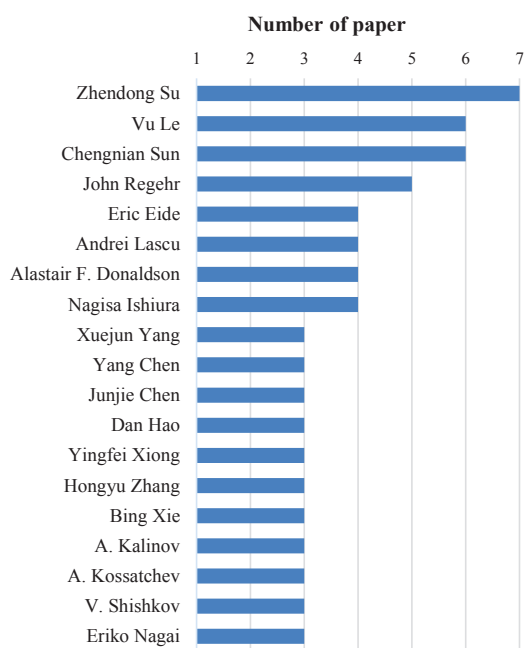}
\caption{The most productive authors}
%\label{fig:fragment}
\label{fig:9}
\end{figure}

\begin{figure}[htb]
\centering
\includegraphics[scale=0.80]{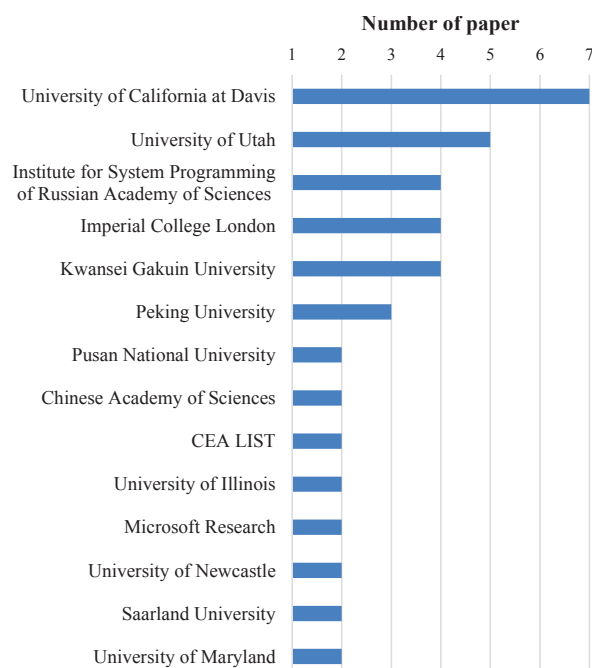}
\caption{The most productive institutions}
%\label{fig:fragment}
\label{fig:10}
\end{figure}

\begin{figure}[htb]
\centering
\includegraphics[scale=0.60]{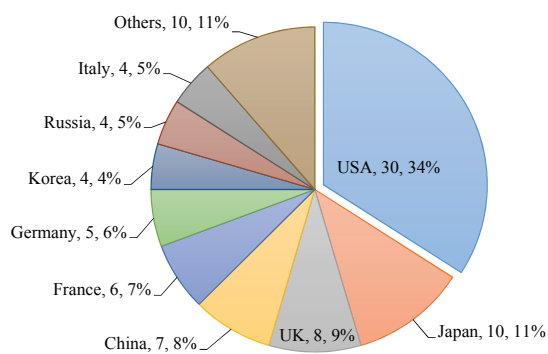}
\caption{The number of papers and ratio for countries/regions}
%\label{fig:fragment}
\label{fig:4}
\end{figure}

\subsubsection{RQ1.1 What are the productive authors, institutions or countries?}
\noindent As for the statistical account of authors and institutions, we list the number of published papers for each author and institution in Fig.~\ref{fig:9} and Fig.~\ref{fig:10}, respectively. The results show that several authors, such as Zhendong Su, Vu Le, and Chengnian Sun, have published more papers related to compiler testing. In addition, most productive authors have collaborations with others. For example, top three productive authors have co-published seven papers in our dataset. Other researchers also make many contributions to promote the development of compiler testing. When calculating the number of papers for each institution, we find that many universities have multiple campuses which are usually located in different areas, and have different research contributions. Thus, we distinguish each campus of a university, and find that the branch campus of University of California at Davis has published the most papers in the compiler testing area.

In addition, we present the number of papers and the ratio of per country/region in Fig.~\ref{fig:4}. The results show that the USA is the leading country with 30 published papers in our dataset, which is consistent with the results obtained in previous studies for ranking analyses of both paper quantity and quality~\cite{Su2010Mapping}. We can also notice that Japan, the UK, China, and France are the most active countries, which indicates that the researchers in these countries tend to pay more attention to compiler testing.

\begin{figure}[htb]
\centering
\includegraphics[scale=0.80]{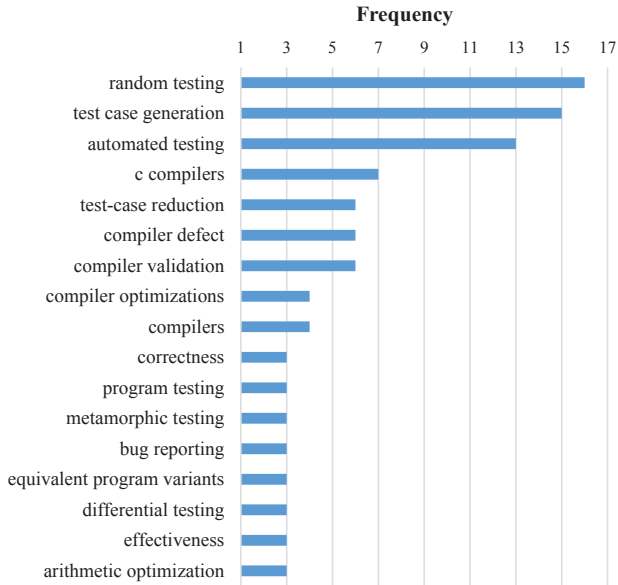}
\caption{The most frequent keywords}
%\label{fig:fragment}
\label{fig:11}
\end{figure}

\begin{figure}[htb]
\centering
\includegraphics[scale=0.60]{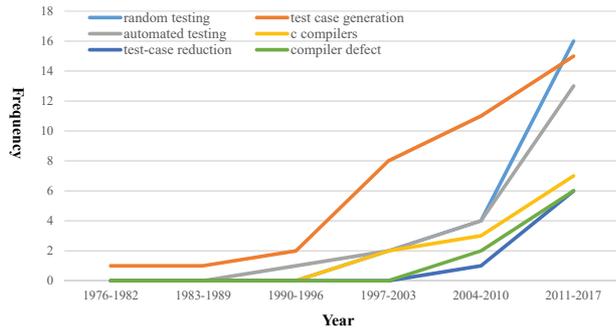}
\caption{The trends of top six keywords}
%\label{fig:fragment}
\label{fig:11-2}
\end{figure}

\subsubsection{RQ1.2 What are the frequent keywords and the trends of popular topics?}

\noindent We also investigate the frequencies of keywords, and list the top-ranked keywords that occur more than three times in Fig.~\ref{fig:11}. The top three active keywords are ``random testing'', ``test case generation'', and ``automated testing''. The first keyword and the third keyword focus on mitigating the test oracle problem, and the second keyword aims to address the difficulty in test case generation.

In addition, we analyze the trends of several keywords. As the papers in our dataset are published from 1976 to 2017, we split the papers into six periods. We accumulate the frequencies of top six keywords on each period to analyze the trends of them, as shown in Fig.~\ref{fig:11-2}. We can observe that the keyword ``test case generation'' has a sharp increase during 1997 and 2003, while presenting a smooth increase after 2003. Indeed, test case generation is a difficult task in compiler testing, and the existing generator tools are only prepared for several languages. In the future, there would be more generator tools to improve testing compilers that support various languages. Notably, the keyword ``random testing'' attracts much more attention after 2010, and becomes the most popular keyword in recent years. Simultaneously, the keyword ``automatic testing'' also shows a sharp increase during the past decade. The reason may be that as an automatic test case generation tool, CSmith, is proposed in 2011, enormous test cases are generated, making the random testing and automatic testing possible. Furthermore, as the compiler is generally of complex structure and the functionality of generating target machine code is the only concern, randomly automatic testing based on enormous test cases is critical in comprehensive testing compilers~\cite{meihong2018can}. Other keywords also show a great increase after 2010, such as test-case reduction, which is an emerging topic in recent years.

\begin{table}[h]
\begin{footnotesize}
\caption{Top scores of ACS}
\label{table:5}
%\label{table:features}
\centering
%\scalebox{0.55}
\begin{tabular}{p{0.5cm}p{3.5cm}p{1.5cm}}
  \hline\noalign{\smallskip}
  no. & author name & score\\
  \noalign{\smallskip} \hline
  1 & John Regehr & 23.00 \\
  2 & Eric Eide & 22.07 \\
  3 & Yang Chen & 18.47 \\
  4 & Xuejun Yang & 17.66 \\
  5 & Zhendong Su & 14.14 \\
  6 & Vu Le & 13.81 \\
  7 & William M. McKeeman & 10.40 \\
  8 & Robert Mandl & 9.97 \\
  9 & Mehrdad Afshari & 8.00 \\
  10 & Chengnian Sun & 6.14 \\
  \hline
\end{tabular}
\end{footnotesize}
%}
\end{table}

\subsubsection{RQ1.3 What are the influential authors and papers in the area of compiler testing?}

\noindent As for the impact of authors, we calculate each ACS score, and list the top ten authors in Table~\ref{table:5}. We can observe that John Regehr, Eric Eide, Yang Chen, Xuejun Yang, and Zhendong Su have higher ACS scores, which indicates that they are excellent researchers in the compiler testing area. Furthermore, we also investigate that most of these authors have published more than four papers in our dataset, such as Zhendong Su, Ve Le, John Regehr, and Eric Eide, which implies that authors with more published papers tend to have higher impact in the compiler testing area.

As for the impact of papers, we calculate the NCII score of each paper, and list the top ten influential papers in Table~\ref{table:7}. From the table, we can see that the most influential paper focuses on addressing the difficulty in test case generation, and creates a tool, Csmith, which can detect many unknown compiler bugs. We can also observe that most papers in top ten are published in the last decade, whereas only two papers are published in the period of eighties and nineties. As for the two early papers, one paper published in 1998 is the first time to propose differential testing technology to test C compilers, and emphasize the importance of avoiding undefined behaviors when generating C test programs, which attracts many following researches. The other paper published in 1985 designs an algebraic method for testing Ada compiler, which is widely-used to generate optimal test cases in software testing.

\textbf{Answer to RQ1:} By conducting the productivity analysis and the impact analysis, we find that the USA is the most influential country with a lot of excellent researchers and institutions in the compiler testing area. The keywords ``random testing'' and ``automated testing'' show a sharp increase in recent years and tend to be the most popular keywords from academia.

\begin{sidewaystable*}
\begin{footnotesize}
\caption{The most influential papers}
\label{table:7}
%\label{table:features}
\centering
%\scalebox{0.55}{
\begin{tabular}{p{0.1cm}<{\centering}p{4.5cm}p{0.4cm}p{0.5cm}p{0.5cm}p{15cm}}
  \hline\noalign{\smallskip}
  no. & title & year & citation & NCII & main contributions\\
  \noalign{\smallskip} \hline
  1 & finding and understanding bugs in c compilers & 2011 & 393 & 56.14 &
  \begin{itemize}
   \item A state-of-the-art test case generator named Csmith is developed, and many previous unknown compiler bugs are found.
   \item A qualitative and quantitative analysis is conducted to characterize the bugs.
  \end{itemize}\\
  2& compiler validation via equivalence modulo inputs & 2014 & 96 & 24.00 &
  \begin{itemize}
   \item A novel testing technology of equivalence module input (EMI) is introduced.
   \item An instance of EMI named Orion is developed for testing C compilers.
   \item A large number of bugs in GCC and LLVM are reported by the evaluation of Orion.
  \end{itemize} \\
  3& test-case reduction for c compiler bugs & 2012 & 97 & 16.17 &
  \begin{itemize}
   \item Three new, domain-specific test case reducers for C codes are proposed in a general framework.
   \item A crucial test-case validity problem is identified, and can be solved by various solutions.
   \item The best reducer is much more effective which produces test cases more than 25 times smaller than that produced by a delta debugger.
  \end{itemize} \\
  4 & taming compiler fuzzers & 2013 & 61 & 12.20 &
  \begin{itemize}
   \item The paper frames the fuzzer taming problem, which has not been addressed by researchers.
   \item The paper exploits the observation that automatic triaging of test cases and automatic test case reduction can be synergistic in accelerating compiler testing.
   \item The paper leverages diverse sources of information about bug-triggering test cases to rank test cases.
   \item The furthest point first (FPF) technology is both faster and more effective to cluster test cases than other clustering algorithms.
   \item Many bugs in a JavaScript engine and a C compiler are found during the fuzzing run.
  \end{itemize} \\
  5 & differential testing for software & 1998 & 208 & 10.40 &
  \begin{itemize}
   \item A new testing technology, differential testing, is proposed, and discovers new bugs in C compilers.
  \end{itemize} \\
  6 & many-core compiler fuzzing & 2015 & 31 & 10.33 &
  \begin{itemize}
   \item The paper provides the evidence on the effectiveness of random differential testing and EMI testing in a new application domain.
   \item The paper proposes three novel methods for generating OpenCL kernels.
   \item An injection of dead-by-construction code enable EMI testing in the context of OpenCL.
   \item More than 50 OpenCL compiler bugs existing in commercial implementations are reported.
  \end{itemize}  \\
  7 & orthogonal latin squares: an application of experiment design to compiler testing & 1985 & 327 & 9.91 &
  \begin{itemize}
   \item The paper proposes a new method for testing compilers, i.e., orthogonal latin squares, which can facilitate exhaustive testing at a fraction of the cost.
   \item The method is effective in designing some tests by using Ada Compiler Validation Capability test suites.
  \end{itemize}  \\
  8 & compiler testing via a theory of sound optimisations in the c11/c++11 memory model & 2013 & 40 & 8.00 &
  \begin{itemize}
   \item A theory of sound optimizations in the C11/C++ memory model is proposed, which covers most optimizations in real compilers.
   \item A bug-hinting tool, cmmtest, is built based on the theory, and discovers some subtle concurrency bugs and unexpected bugs in the C11/C++ memory model.
  \end{itemize} \\
  9 & testing an optimising compiler by generating random lambda terms & 2011 & 53 & 7.57 &
  \begin{itemize}
   \item The paper provides a workable solution to generate random and type-correct lambda terms, and discovers many bugs in the Glasgow Haskell compiler.
  \end{itemize} \\
  10 & volatiles are miscompiled, and what to do about it & 2008 & 72 & 7.20 &
  \begin{itemize}
   \item The paper shows that C's volatile qualifier in compilers can produce incorrect object codes.
   \item The paper proposes a technique for generating C programs randomly.
   \item A new testing technique, access summary testing, is proposed, which is effective and automatical at detecting compiler bugs.
   \item The paper shows that the impact of compiler bugs can be mitigated by introducing small helper functions into a program.
   \item Several recommendations are provided for application developers and compiler developers.
  \end{itemize} \\
  \hline
\end{tabular}
\end{footnotesize}
%}
\end{sidewaystable*}

\subsection{Investigation to RQ2}
\label{subsec:investigationRQ2}
\noindent We detect the most frequently tested compilers, popular compiler testing technologies, and available tools by conducting the content analysis.

\begin{figure*}[!ht]
\centering
\includegraphics[scale=0.90]{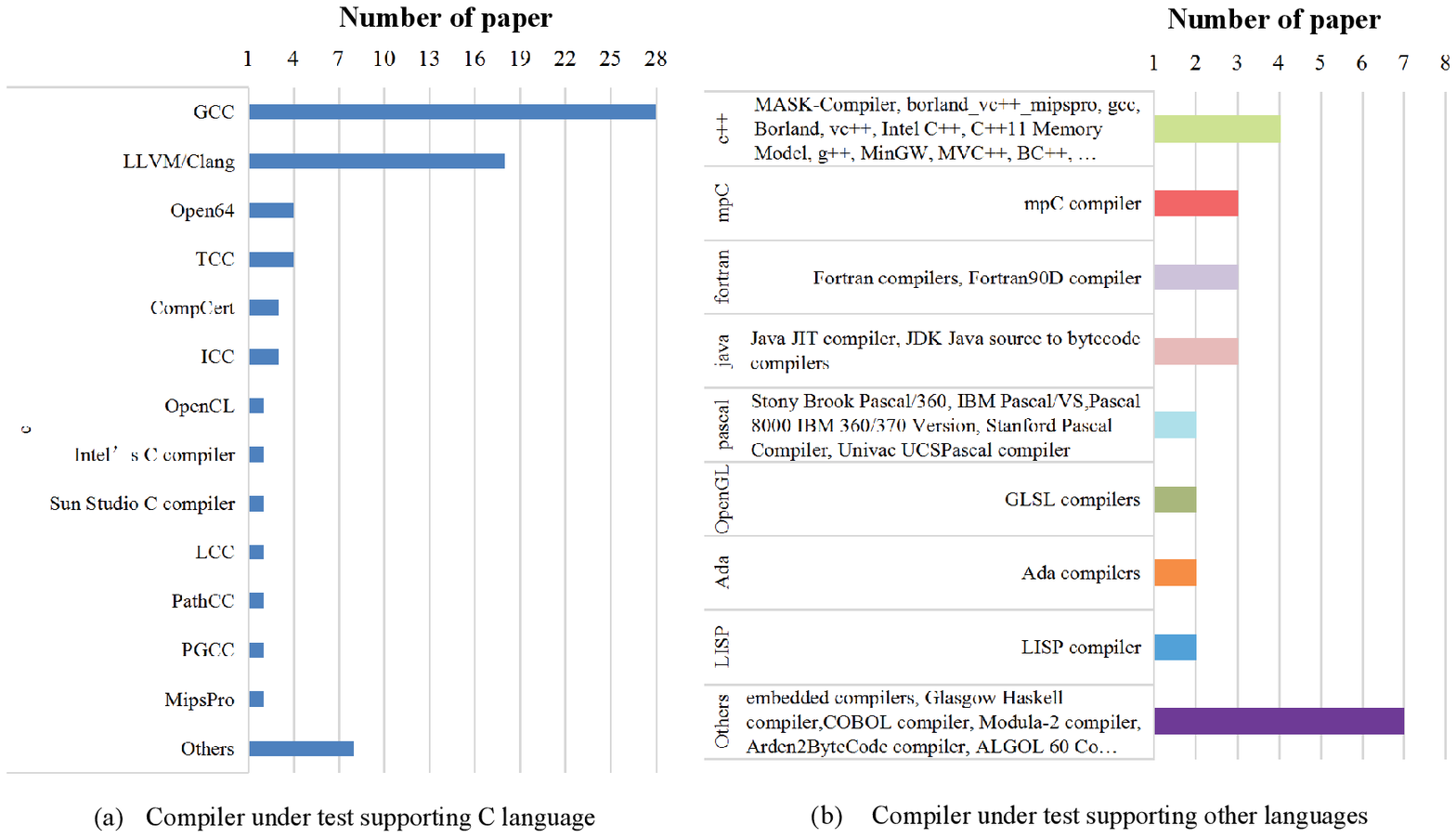}
\caption{Compiler under test supporting different languages}
\label{fig:12}
%\label{fig:fscore}
\end{figure*}

\subsubsection{RQ2.1 What compilers are frequently tested?}

\noindent We calculate the frequencies of tested compilers used by researchers, and list the number of papers of tested compilers in Fig.~\ref{fig:12}. Notably, most papers use various types of compilers that support the same language or one type of compiler with different optimization levels. The results show that C compilers are frequently tested by most papers, especially GCC and LLVM/Clang. In fact, GCC~\cite{sun2016toward} is a compiler system supporting various languages and target architectures. LLVM~\cite{lattner2004llvm} is another popular compiler infrastructure, and has drawn much attention from academia. Other compilers supporting different languages also attract researchers to test their correctness, ranging from C++, Java to Pascal. As a result, the quality of compilers is critical for any language, and the compilers should be comprehensively tested.

\subsubsection{RQ2.2 What test cases and testing technologies are employed when testing compilers?}

\begin{figure}[htb]
\centering
\includegraphics[scale=0.70]{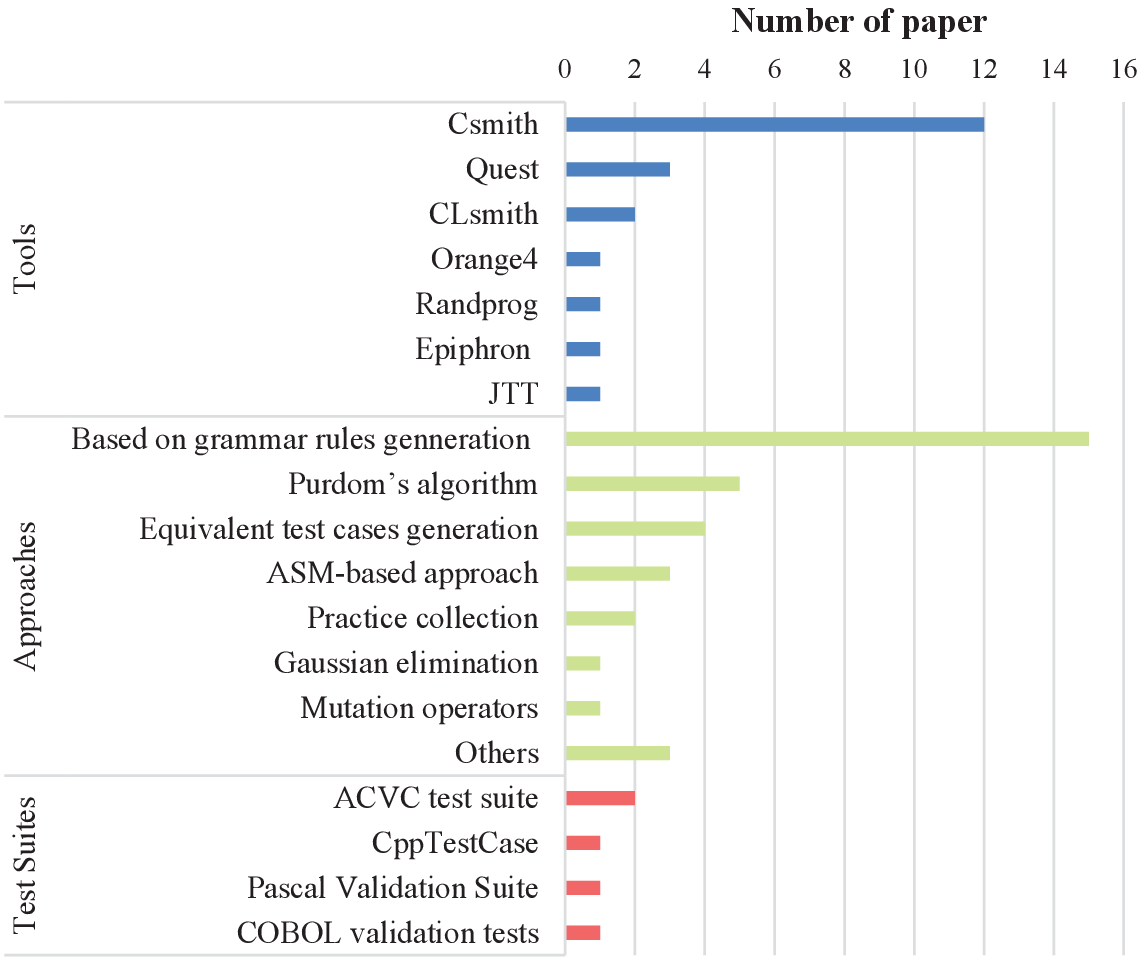}
\caption{Test case generation tool/approach}
%\label{fig:fragment}
\label{fig:13}
\end{figure}

\noindent In order to test compilers, test cases are needed as the inputs of compilers (see Section 3.2). To effectively generate abundant test cases that conform to language standards and specifications, many approaches and tools are proposed to generate random test cases without undefined behaviors. We identify each test case generator, and list the frequencies of these generators in Fig.~\ref{fig:13}. The results show that most test cases are generated based on the language grammar rules and the coverage criteria. Several tools are frequently adopted by researchers, such as Csmith, Quest, CLsmith, Orange4, randprog, Epiphron, and JTT, whereas test suits are rarely employed because of the limitation of definite test cases. As for the automatic test case generators, Csmith is the most widely-used tool for C language test case generation, because Csmith covers a broad range of syntax of the C language, including arrays, structs, conditional statements, loop statements, and function calls, which is more expressive than other tools. In addition, several approaches can also generate abundant test cases. For example, Purdom' algorithm is a popular approach which generates test cases based on the language specifications, and has been extended by other test case generation approaches. Other approaches based on metamorphic testing and SPE can generate a set of semantic equivalence test cases as the inputs of compilers.

\begin{figure}[htb]
\centering
\includegraphics[scale=0.80]{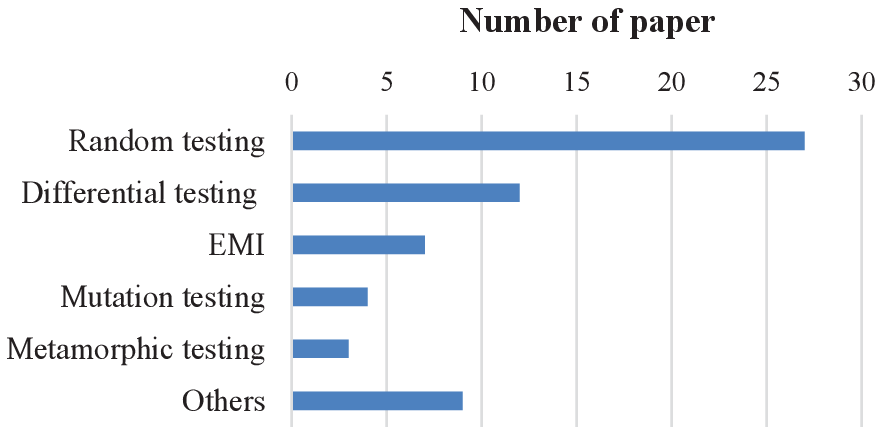}
\caption{Different compiler testing technologies}
%\label{fig:fragment}
\label{fig:14}
\end{figure}

After the test cases are fed into compilers, we need to employ testing technology to test compilers based on these test cases (see Section 3.3). Different testing technologies are proposed to guarantee the quality of compilers as shown in Fig.~\ref{fig:14}. Random testing and differential testing are the two most frequently used technologies. Both of these two technologies can test compilers using randomly generated test cases as long as time allows. The view behind the differential testing is that if more than two compilers under the same test cases produce different results, there is a bug in at least one compiler. EMI derived from metamorphic testing attempts to construct equivalence-preservation relations to generate equivalent test cases for testing compilers. In addition, EMI is simple and widely applicable, which has been employed by many researchers. Mutation testing, as a trade off between the efficiency and the effectiveness in compiler testing, detects a mutant if errors manifest in a mutant,  which is adopted by several researchers. Other testing approaches, such as Optimizer Testing Kit approach~\cite{Zelenov2012Model} and SPE approach, are also employed by researchers to test different compilers, and show their own effectiveness on bug detection.

\begin{figure}[htb]
\centering
\includegraphics[scale=0.80]{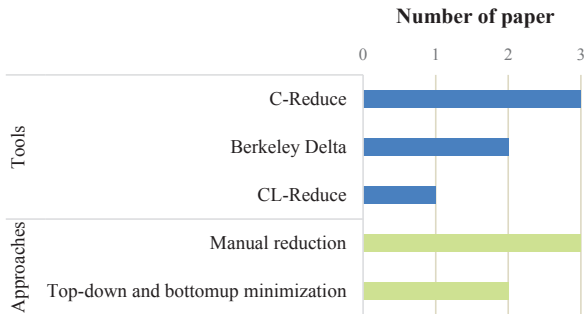}
\caption{Tools of test case reducer}
%\label{fig:fragment}
\label{fig:15}
\end{figure}

\subsubsection{RQ2.3 How to reduce the large test cases before reporting?}

\noindent Once a test case triggers a bug, the test case should be reduced before reporting, because large test cases are tedious and time-consuming for developers to find the root cause of the bug (see Section 3.4). Several automatic reducers are developed to help reduce large test cases, and ensure that the reduced test cases still trigger the same bug, and do not introduce new undefined behaviors. We identify each reducer employed by researchers, and list the frequencies of these reducers in Fig.~\ref{fig:15}. C-reduce is the most popular reducer due to the high efficiency and effectiveness on reducing test cases. Berkeley Delta and CL-Reduce are also adopted by researchers when there is a need to reduce the large size of test cases. Another reduction approach~\cite{Nagai2012Random} employs top-down minimization and bottom-up minimization algorithms to reduce the arithmetic expressions to a small program.

\textbf{Answer to RQ2:} By conducting the content analysis, we find that C compilers are frequently tested by academia, and random testing is the most popular testing technology. In addition, several tools for test case generation and reduction are available for the public, such as Csmith and C-reduce. However, the number of papers implementing test case reduction is much smaller than that of test case generation, which also encourages researchers pay more attention on the large test case reduction, even for the real world projects.

\begin{figure*}[!ht]
\centering
\includegraphics[scale=0.70]{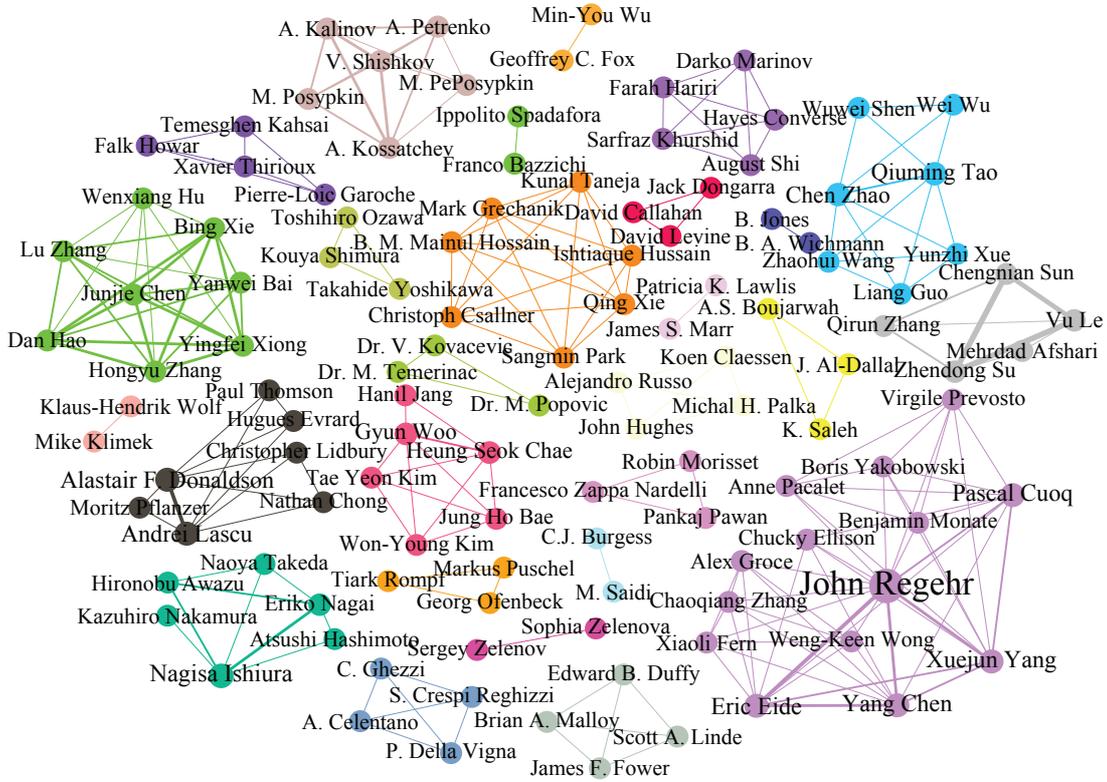}
\caption{Co-authorship network}
%\label{fig:fragment}
\label{fig:5}
\end{figure*}

\subsection{Investigation to RQ3}
\label{subsec:investigationRQ3}
\noindent We explore the relationships among authors and their interests by two collaboration networks using association rules, i.e., the co-authorship network and the author co-keyword network. In addition, we analyze the relationships between the frequent co-occurrence keywords by the keyword co-occurrence network. In the following subsections, we present and analyze the collaborations in these networks.

\subsubsection{RQ3.1 What are the relationships among authors of compiler testing?}

\noindent We present the collaboration relationships among authors in the co-authorship network. Actually, the number of authors in this network is affected by the minimal support \textit{$t_{s}$} and the minimal confidence \textit{$t_{c}$}. When the minimal \textit{$t_{s}$} is set to 0.017, and the minimal \textit{$t_{c}$} is set to 0, all the frequent pairs of authors will be included in the network. Thus, the co-authorship network has the maximum number of authors in the compiler testing area.

The co-authorship network is shown in Fig.~\ref{fig:5} which contains 119 authors and 229 links, and includes 27 authorship communities with the modularity of 0.893. In Fig.~\ref{fig:5}, authors in compiler testing distribute in several scattered communities, which only 32 authors (27\% rate) collaborate with more than five authors, and only one author (0.8\% rate) collaborates with more than ten authors. The strength of collaborations is in a small rang from one to six cooperative times, and only five pairs of authors (2.18\% rate) collaborate with each other more than three times. These communities are isolated from each other, among which seven communities follow an edge structure, six communities follow a triangle structure, four communities follow a quadrilateral structure, while the other ten communities follow a complex network structure. In the following discussions, we only discuss the ten complex communities, and use the high degree author or the productive author to represent a community, such as the John Regehr community and the Junjie Chen community.

In Fig.~\ref{fig:5}, the John Regehr community is the most complex community with 14 collaborators and 51 links. In this community, John Regehr is the central author, and has other 13 collaborators, especially collaborating with Eric Eide for four times. John Regehr is also a productive author in Fig.~\ref{fig:9}, and receives the most highest ACS score in Table~\ref{table:5}. Other productive authors also appear in this community, such as Yang Chen and Xuejun Yang. All of these authors have broad collaborations that form the biggest community in the compiler testing area. In the Junjie Chen community and the Ishtiaque Hussain community, there are eight and seven collaborators respectively. Each pair of the collaborators in these two communities has co-authored papers in our dataset. Specifically, there are three co-authored papers related to compiler testing in the Junjie Chen community as shown in Fig.~\ref{fig:9}.

The other seven complex communities have more than five collaborators, and surround with several productive authors, such as Zhendong Su and Alastair F. Donaldson. Although there are only five authors in the Zhendong Su community, all the authors dominate the state-of-the-art technologies on compiler testing. Another two communities, i.e., the Alastair F. Donaldson community and the Nagisa Ishiura community, have several strong associations among collaborators, while other four complex communities have more collaborators but weak associations.

We also investigate the determining factors of the co-authorship phenomenon, and the impact of papers affected by the collaborations. In our dataset, the authors of more than half of papers are from the same institution, and of more than two thirds papers are from the same country. For examples, all the authors in the Zhendong Su community are from the University of California at Davis, and all the authors in the Junjie Chen community are from China. Furthermore, collaborations can increase the number of papers, as well as the accepted rate for publication in a conference or journal. However, the quality of co-authored work is the most critical factor on the acception for a top conference or journal, which can greatly improve the influence of a paper. In fact, collaborations can certainly improve the quality of work, especially collaborating with some productive authors, such as the most influential papers listed in Table~\ref{table:7}, which has received much more citation numbers from academia.

\begin{figure}[htb]
\centering
\includegraphics[scale=0.35]{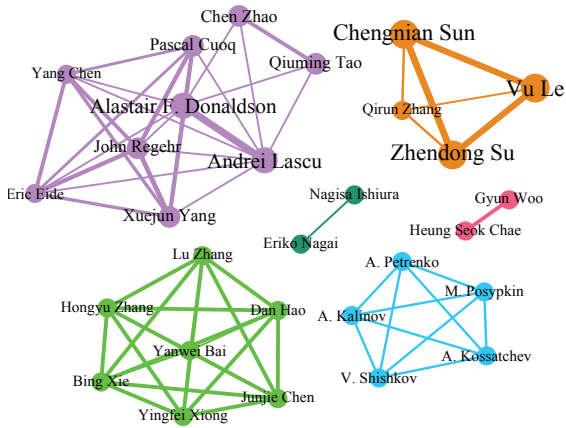}
\caption{Author co-keyword network}
%\label{fig:fragment}
\label{fig:6}
\end{figure}

\subsubsection{RQ3.2 What are the same interests of authors?}

\noindent In this subsection, we analyze the interests among authors using the author co-keyword network. Similarly, given the minimal support \textit{$t_{s}$} of 0.03, if the minimal confidence \textit{$t_{c}$} is set to 0.09, authors that published more than two papers using the same keywords more than three times are included in this network. We set this pair of parameters in association rules because it can significantly detect the same interests among productive authors, and present a clear topological structure in the network.

The author co-keyword network is shown in Fig.~\ref{fig:6}. We can see that there are six communities clustered by 29 authors and 65 links. Each community is composed of authors with the same topic because the authors in a community tend to use the same keywords. Thus, different communities share different topics in the compiler testing area. As the same in co-authorship network, we use the high degree author or the productive author to represent a community, such as the Alastair F. Donaldson community and the Junjie Chen community.

In Fig.~\ref{fig:6}, six communities are isolated from others. Among these communities, the largest community is dominated by Alastair F. Donaldson, who is interested in graphics shader compilers and many-core compiler testing, and focuses on CLsmith and CL-reduce tools for test case generation and reduction. Furthermore, this community is composed of three collaboration communities in co-authorship network, namely the John Regehr community, the Alastair F. Donaldson community and the Qiuming Tao community, as shown in Fig.~\ref{fig:5}. The Zhendong Su community aims at issues of test case generation and compiler testing technology, such as SPE and EMI.

The other three isolated communities have the same collaboration communities as Fig.~\ref{fig:5} shows. The Junjie Chen community aims to prioritize test cases for compilers to accelerate the process of compiler testing~\cite{Chen2016Test,Chen2017Learning}. The A. Kalinov community focuses on test case generation for mpC compiler~\cite{Kalinov2002Using}, and the Heung Seok Chae community is interested in test case reduction for retargeted compilers~\cite{Chae2011An,Woo2007An}.

\begin{figure}[htb]
\centering
\includegraphics[scale=0.40]{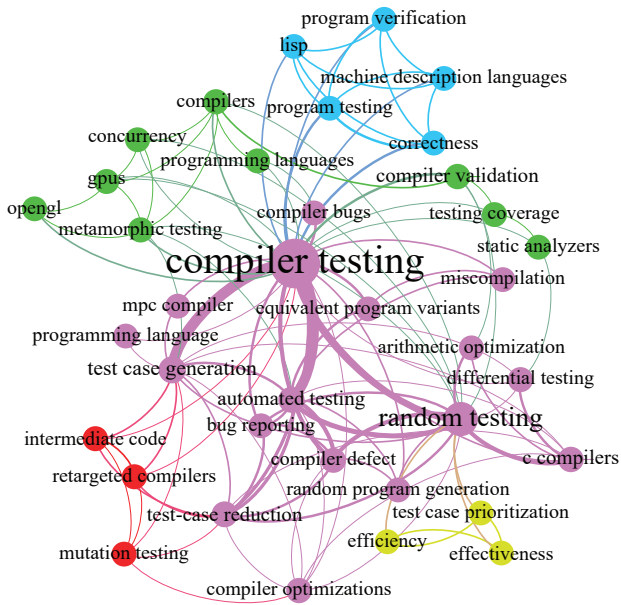}
\caption{Author co-keyword network}
%\label{fig:fragment}
\label{fig:7}
\end{figure}

\subsubsection{RQ3.3 What are the frequent co-occurrence keywords in the area of compiler testing?}

\noindent In this subsection, we present the major topics and the links between keywords by the keyword co-occurrence network. We analyze the structure of this network mined at the minimal support \textit{$t_{s}$} of 0.033. When the minimal confidence \textit{$t_{c}$} is set to 0, all the keywords that occurred more than two times are included in this network. We select these parameters due to two reasons. First, 33.64\% frequent keywords can be included in this network when the minimal set is 0.033. Second, we can discover a clear topological structure among these frequent keywords.

The keyword co-occurrence network is shown in Fig.~\ref{fig:7} which consists of 37 keywords and 117 links, and forms five communities with the modularity of 0.248. We use the high degree node or the major topic to define a community. Furthermore, to avoid the ambiguity with nodes in a community, we use the first-word-capitalized name to refer to a community, such as the Compiler Test community.

In Fig.~\ref{fig:7}, most communities have complex links with each other. We can see that a central keyword ``compiler testing'' is linked with most keywords in these communities. In addition, two keywords ``random testing'' and ``test case generation'' have strong associations with the central keyword, indicating that these two keywords are the most important topics related to compiler testing, and attract more researchers to focus on. Simultaneously, these three core keywords dominate the largest community in this network which we define as the Compiler Test community. In this community, there are 17 keywords, including the issues and the solutions related to compiler testing, such as ``test-case reduction'', ``equivalent program variants'', ``random program generation'', and ``differential testing'', which defines a fine-grained category of compiler testing. Another three keywords, namely ``efficiency'', ``effectiveness'', and ``test case prioritization'', also surround with a high degree keyword ``random testing'', which form a community aimed at improving the performance of compiler testing technologies.

The Compiler Validation community is also a relatively larger community that contains nine keywords which focuses on test case generation for many core compilers. This community surrounds with keywords ``gpus'', ``opengl'', ``concurrency'', ``testing coverage'', ``static analyzers'', ``compiler validation'', ``programming languages'', ``compilers'', and ``metamorphic testing''. However, two other communities are relatively smaller. The keywords in these two communities are linked with each other which referred as the Program Verification community and the Retargeted Compiler community, respectively.

\textbf{Answer to RQ3:} With the analysis of the co-authorship network and the author co-keyword network, we find that the co-authorship in the compiler testing area distributes in several scattered communities. Authors in the same institution and the same country tend to collaborate with each other. In addition, most productive authors have broad interests, and the collaborations with the productive authors can improve the influences of papers to a certain extent. By constructing the keyword co-occurrence network, we find that the test case generation and the test oracle problem are the two most critical issues in compiler testing, which surround with abundant relevant keywords.

\section{Related work}
\label{sec:related}
\noindent The most relevant work is literature analysis. In this section, the majority of related work can be classified into two aspects, i.e, the bibliometric analysis and the collaboration analysis.

\subsection{Bibliometric analysis}
\label{subsec:bibliometrics}
\noindent A large number of bibliometric studies have been published in software engineering. Wohlin et al.~\cite{Wohlin2005An,Wohlin2007An,Wohlin2008An,wohlin2009an} analyzed the highly cited papers in software engineering published from 1999 to 2002. Wong et al.~\cite{wong2008assessment,wong2009assessment,wong2011assessment} identified top-15 researchers and institutions for two five-year periods between 2008 and 2011. The rankings were based on the number of published papers from seven leading software engineering journals.

Focusing on the sub-areas of software engineering, Souza et al.~\cite{Souza2011Ten} presented a bibliometric analysis for ten years of search-based software engineering that covered 740 papers from 2001 to 2010. Jiang et al.~\cite{He2016ming} constructed a publication analysis framework to present some important domain knowledge for mining software repositories. Some recent systematic mapping studies also included bibliometric analysis of sub-areas of software engineering, e.g., web application testing~\cite{garousi2013a}.

In previous work, Garousi et al.~\cite{VAHID2013A} conducted the first quantitative bibliometric analysis in total about 60\% of the software engineering literature, and reported interesting findings, such as the USA is the clear leader, but the contributions to software engineering by the American researchers have decreased from 71.43\% (in 1980) to 14.90\% (in 2008). More recently, Garousi et al.~\cite{Garousi2016Highly} utilized automated topic analysis to characterize and understand massive software engineering literature.

\subsection{Collaboration analysis}
\label{subsec:Collaboration}

\noindent The co-authorship network aims to find the cooperative relationship among authors. Velden et al.~\cite{Velden2010A} studied patterns of collaboration in the co-authorship networks with the data obtained from Web of Science. They identified two types of coauthor-linking patterns between authorship communities with the name disambiguation. But they distorted the topological structure of the co-authorship networks in some cases because a small set of common surnames are widely used in some East Asian countries. Madaan et al.~\cite{Madaan2015Evolution} found interesting features in the co-authorship network, such as the collaborations between researchers is increasing over time, and few researchers published a large number of papers in DBLP Computer Science Bibliographic database.

Su et al.~\cite{Su2010Mapping} created a three-dimensional research, focusing on a parallel network, i.e., the keyword co-occurrence network, and a two-dimensional knowledge map to visualize the knowledge structure using the data of journal papers.

The difference between our work and previous work is that we employ both bibliometric and collaboration analyses for compiler testing literature analysis. In the bibliometric analysis, we not only distinguish the authors¡¯ names with their institutions to avoid ambiguities, but also combine the ACS score and the NCII score to measure the impact of authors. In the collaboration analysis, we first incorporate the social network and the data mining technique to construct the co-authorship network, the author co-keyword network, and the keyword co-occurrence network to help mine useful collaborations.

\section{Conclusion \& Future work}
\label{sec:conc}
\noindent In this study, we present a literature analysis framework, to comprehensively characterize and understand the compiler testing area. We illustrate how each component works in the framework and obtain some useful information after conducting each component. The major contributions of this paper include two aspects. In the aspect of bibliometics analysis, we find that the USA dominates the area of compiler testing, having a large number of influential researchers, such as Zhendong Su, Vu Le, and Chengnian Sun. The keyword ``random testing'' is the most frequently used keyword by researchers, and C compilers are the most frequently tested compilers. In the aspect of collaboration analysis, we construct three collaboration networks, and find that collaborations with productive researchers can improve both the accepted rate and the quality of papers in the co-authorship network. In addition, we detect several researchers with the same interests in the author co-keyword network, and some fine-grained categories of compiler testing in the keyword co-occurrence network.

Although the previous work has proposed various solutions to the issues existing in the compiler testing area, there still remain several interesting challenges that need to be addressed in the future, such as using the real-world projects to test compilers, reducing test cases for multiple files, improving both the effectiveness and efficiency of compiler testing technologies, etc. In the future, we will focus on these challenges to improve test compilers and hope more researchers devote to compiler testing to boom this area.

\end{document}